\documentclass[12pt, onecolumn, draftclsnofoot]{IEEEtran}
\usepackage{bbding}
\usepackage{url}
\usepackage{booktabs}
\usepackage{multirow}
\usepackage{fancybox}
\usepackage[dvipsnames,svgnames,x11names]{xcolor}
\usepackage{graphicx}
\usepackage{bbm}
\usepackage{cite}
\usepackage{epsfig, psfrag, amsmath, amssymb, amsfonts, latexsym, eucal, balance, hyperref,indentfirst,bookmark, amsthm}

\usepackage{subfig}

\graphicspath{{figs/}}
\begin{document}

\title{Towards a Tractable Delay Analysis in Ultradense Networks}
\author{Yi Zhong, \emph{Member, IEEE}, Martin Haenggi, \emph{Fellow, IEEE}, \\Fu-Chun Zheng, \emph{Senior Member, IEEE}, Wenyi Zhang, \emph{Senior Member, IEEE}, \\
Tony Q.S. Quek, \emph{Fellow, IEEE}, Weili Nie
\thanks{
Yi Zhong is with Huazhong
University of Science and Technology, Wuhan, P. R. China (e-mail: yzhong@hust.edu.cn).
Martin Haenggi is with University of Notre Dame, Notre Dame 46556, USA (email: mhaenggi@nd.edu).
Fu-Chun Zheng is with Harbin Institute of Technology, Shenzhen, China and the University of York, York, UK (Email: fzheng@ieee.org).
Wenyi Zhang is with University of Science and Technology of China, Hefei 230027, China (e-mail: wenyizha@ustc.edu.cn).
Tony Q.S. Quek is with Singapore University of Technology and Design, Singapore (e-mail: tonyquek@sutd.edu.sg).
Weili Nie was with the National Mobile Communications Research Laboratory, Southeast University, Nanjing, China, and is now with the Department of Electrical and Computer Engineering, Rice University, Houston, TX, USA (e-mail: wn8@rice.edu).
}
}

\maketitle

%
\begin{abstract}
Meeting the diverse delay requirements of emerging wireless applications is one of the most critical goals for the design of ultradense networks.
Though the delay of point-to-point communications has been well investigated using classical queueing theory, the delay of multi-point to multi-point communications, such as in ultradense networks, has not been explored in depth.
The main technical difficulty lies in the interacting queues problem, in which the service rate is coupled with the statuses of other queues.
In this article, we elaborate on the main challenges in the delay analysis in ultradense networks.
Several promising approaches, such as introducing the dominant system and the simplified system, to bypass these difficulties are proposed and summarized to provide useful guidance.
\end{abstract}

\newpage

\section*{Need for New Delay Analysis}
The emergence of new latency-critical applications, such as intelligent manufacturing, remote control, auxiliary driving, and automatic driving, has led to a variety of delay requirements in wireless networks.
Specifically, the end-to-end delay for 5G is required to be less than $10$ms (about $1/10$ of the delay requirement for 4G), and for some special applications such as the tactile Internet \cite{fettweis2014tactile}, the delay is required to be less than $1$ms.
Ultradense networking is a promising architecture to meet the delay requirements of the 5G wireless networks \cite{ge20165g,7811838,7961156}.

A theoretic analysis of the delay in ultradense networks is imperative to guide the practice.
However, such delay is difficult to calculate since it is an intricate function of all links and is affected by a variety of factors such as network load, medium access control (MAC), path loss, and so on.
In general, these factors can be classified into three aspects:
\begin{itemize}
\item Random arrival and queueing of packets at the terminals;
\item Spatial models for transmitters and receivers, which affect the path loss;
\item Channel fluctuations and transmission mechanisms, which affect the delivery of packets.
\end{itemize}

Existing works on the delay analysis in networks typically focus on one or at most two of these aspects:
analyses using queueing theory mainly evaluate scheduling algorithms but usually ignore the interference and noise \cite{anantharam1991stability};
analyses based on stochastic geometry often ignore the queueing process and focus on the reliability or throughput in backlogged networks \cite{baccelli2009stochastic};
analyses based on multiuser information theory usually evaluate the network capacity  \cite{andrews2008rethinking}.
In order to accurately characterize the delay in ultradense networks, all aspects should be considered. Such a combination of all these methods, however, is long known to be notoriously difficult \cite{6204337}.

This article first reviews the treatment of delay in classical queueing theory and describes the interacting queues problem.
Then, the fundamental challenges that make the delay analysis difficult in ultradense networks are discussed.
In order to handle these difficulties, several promising approaches are proposed and evaluated.

\subsection*{End-to-end Delay}
Delay in this article refers to the end-to-end delay, which is the duration between generating a packet at the transmitter and successfully decoding it at the receiver.
Note that the end-to-end delay here is the delay within the wireless access network, while
the delay in the core wired networks is beyond the scope of this article.
Generally, the end-to-end delay consists of the processing delay, the queueing delay, the transmission delay, and the propagation delay.
The processing delay is the time it takes to generate the packets, which is about several microseconds.
The queueing delay is the waiting time of a packet until it is served.
The transmission delay is the time to successfully transmit a packet that is served.
When a retransmission mechanism is applied, the delay caused by waiting and retransmission is included in the transmission delay.
The propagation delay is the duration between a packet leaving the transmitter and reaching the receiver, which is calculated by dividing transmission distance by the speed of electromagnetic waves.
In the multihop case, these four delay elements apply to each link.
In ultradense networks, the processing delay and the propagation delay are negligible compared to the queueing delay and the transmission delay, which are the focus of this article.

\subsection*{Classical Queueing Theory}
In classical queueing theory, Kendall's notation, A/S/C, is applied to characterize the queueing problems, where A denotes the time intervals between two adjacent arrivals, S describes the service process of packets, and C is the number of servers. 
The delay in classical queueing scenarios with only one server or where the service rates of different queues are independent is well studied.
For instance, in M/M/1 queueing problem (see Fig. 1(a)), where `M' denotes `Markovian', there is a single server, the arrival process of the packets is a Poisson process with arrival rate $\lambda$, the service time of each packet is exponentially distributed with mean $\mu$, and thus the mean delay is $1/(\mu-\lambda)$.
However, the emergence of multiple queues and servers in ultradense networks substantially increases the difficulty of the queueing problem; moreover, the coupling of the service rates of different queues leads to the \emph{interacting queues problem}.

\subsection*{Interacting Queues Problem}
A typical interacting queues problem can be described as follows.
Consider an ALOHA system with the time being divided to discrete slots with the same duration.
Assume that there are $N$ terminals, and the arrival processes are independent for different terminals. Each terminal is active with a certain probability and
delivers its head-of-line packet in each time slot if its queue is nonempty. If there is more than one simultaneous transmissions, a collision occurs, and all involved packets will wait to be retransmitted.
The essential difficulty of the interacting queues problem lies in that
the service rate of each queue depends on the statuses of all other queues.
Figure 1(b) shows an example of the interacting queues problem when there are only two terminals.
If one queue is empty, the corresponding transmitter will not interfere the other link; thus the service rate of the other link increases, and its queue becomes empty more quickly.

The delay for interacting queues is difficult to analyze, and existing work has only explored the stability issue, i.e.,
whether the queues will grow without bounds.
The stability region is the range of arrival rate that guarantees the stability of all queues.
For the above system with $N$ queues, the stability region has been found only for $N=2$ and $N=3$ \cite{rao1988stability}. If the number of queues is finite and more than three, only sufficient or necessary conditions for stability are known.
In addition, the interference in ultradense networks cannot be just modeled as collisions.

\section*{Challenges in The Delay Analysis in Ultradense Networks}
The delay is influenced by the queueing and service processes of packets.
The queueing process with multiple queues is different from the classical queueing problems due to the interacting queues, while the service process is directly determined by the MAC and the signal-to-interference-plus-noise ratio (SINR). The key challenges in the delay analysis are the following:

\subsection*{Challenge 1: Randomness in the Spatial Deployment}
Delay is directly related to the service rate determined by the SINR, which is significantly affected by the inter-node distances.
The distance between transmitter and receiver in the desired link determines the received power for the desired signal.
Moreover, in practice, the ultradense networks are interference-limited, i.e., noise is negligible compared with interference, making  interference a main factor that affects the SINR and, in turn, the delay.
Notice that the sum interference power depends on the distances between interfering transmitters and the desired receiver.
All these link distances are functions of the network geometry.

The spatial structure of the heterogeneous ultradense networks is by no means regular. The irregularity exists even in meticulously deployed macro base stations, due to the restrictions of locations of sites, the irregular spatial distribution of the traffic, and so on. For heterogeneous ultradense networks, the irregularity is more evident since the deployment of the dense access points is less elaborately planned and more likely to appear random.
This kind of spatial irregularity is termed \emph{deployment randomness}.

Traditionally, the spatial distribution of the nodes in cellular wireless networks is modeled by regular grids. For example, the hexagonal grid is used to characterize the cells generated by macro base stations.
However, the regular grid does not capture the deployment randomness in ultradense networks.
Fortunately, a powerful mathematical tool, point process theory, is available to handle the spatial modeling of the deployment randomness \cite{net:Andrews10commag}. Point process theory represents the location of each node as a point in a spatial point process and permits the analytical characterization of a number of network metrics, including coverage probability, mean rate, and area spectral efficiency.


\subsection*{Challenge 2: Quasi-static Deployment}
While point process theory is widely used to model the topology of the wireless networks, the issue of delay has received considerably less attention. A main reason is that the delay is a long-term metric while the coverage probability, the mean achievable rate, etc., are obtained by considering just a snapshot of the network. In order to discuss long-term metrics, the static nature of the ultradense networks, i.e., the fact that the locations of nodes remain unchanged during a relatively long time once they are deployed, needs to be considered. 
Ultradense networks are approximately static since the topology does not change drastically for a short time.
From the receivers' perspective, the locations of the
interferers, determined in the deployment stage, are uncertain and may be considered as random.
Therefore, the static but random locations can be considered as the common randomness over different time slots, leading to temporal interference correlation.
Such static networks are more challenging to analyze than the high-mobility networks (where the topology is regenerated independently in each time slot)
because inherent correlations of signal and interference persist across
different time slots.

\subsection*{Challenge 3: Dynamics in Channel and MAC}
\textbf{Channel Fluctuations} -- The delay depends on the SINR, which, in turn, determines the quality of the wireless channels. The channel gain in a ultradense network is determined by the path loss and the fading.
The path loss is influenced by factors like propagation medium (moist or dry air), link distances, terrain contours, and height of antennas.
The channel fading, categorized as slow fading and fast fading, varies with time, geographical location and propagation environment, and is often modeled as a stochastic process.
Channel fluctuations may result in a loss of signal power and cause poor delay performance.
Accurately modeling the effect of channel fluctuations on the delay is again difficult since a large number of links coexist, each of them experiencing independent or dependent channel fluctuations.

\textbf{MAC} -- The MAC determines how resources (time, space, bandwidth) are allocated to the links. The effect of the MAC on the delay is two-fold: Firstly, it
has a significant influence on the SINR at the receiver.
Due to the MAC, all the transmitters are divided into two sets: the set containing all transmitters using a certain carrier, and the set containing all other transmitters.
Only the set of transmitters using the same carrier at the same moment cause interference.
Secondly, as part of the MAC, different scheduling policies, such as First-In-First-Out (FIFO), round-robin, and proportional fair, lead to a different delay performance.
Though the delay of various scheduling policies in classical queueing theory is well studied, it becomes complicated to analyze in ultradense networks where  scheduling occurs across a large number of queueing nodes, usually in a distributed fashion.

\subsection*{Challenge 4: Interaction among Queues}
The interacting queues problem in ultradense networks is more complicated than that in the aforementioned slotted ALOHA system (see Fig. 1(c)).
The main differences between the two systems can be attributed to the physical layer as well as the MAC layer respectively.

\textbf{Physical Layer} -- In a collision-based slotted ALOHA system, the transmission mechanism is simple: a packet transmission fails if two or more transmitters in the system are scheduled at the same time. However, in ultradense networks, the packet delivery process is not just determined by the busy statuses of all transmitters but directly affected by the aggregated interference from all active links. Due to link adaptation or adaptive modulation and coding, the transmission rate is adjusted adaptively according to the SINR, which is related to the propagation environment of all links. Therefore, the queues in ultradense networks are coupled in a complicated way.

\textbf{MAC Layer} -- MAC protocol in ultradense networks is usually much more sophisticated than ALOHA. For instance, in the downlink of a ultradense network, each base station may serve multiple users, and user scheduling is introduced to guarantee that most users can be served fairly. If one separate queue is maintained for each user at the base station, there are many queues at each base station. In the multi-cell scenario, the interaction exists between queues of the same cell and queues of different cells (intra-cell and inter-cell interaction).

\section*{Promising Approaches}
\subsection*{Network Stability: The First Step}
Before the delay analysis, a key issue is to explore the stability of the queues in ultradense networks.
Loynes' theorem \cite{loynes1962stability} indicates that
the sufficient and necessary condition for the stability of a single queue is that the arrival rate is less than the service rate.
However, for the ultradense networks modeled by Poisson Point Processes (PPPs), the strict stability (all queues in the networks are stable) cannot be achieved since
two nodes can be arbitrary near to each other, and the distance between nearby transmissions can be arbitrary small.
Therefore, there always exist some links that experience strong interference and, consequently, their queues are unstable even with arbitrarily small arrival rate.
A weaker form of stability is $\varepsilon$-stability, which indicates that the proportion of unstable queues among all queues is at most a predefined value $\varepsilon$.
Assume that the arrival rate of all queues in the network is the same. Then, there exists a critical arrival rate \cite{7486114}: if the practical arrival rate is smaller than the critical arrival rate, the network will be $\varepsilon$-stable; otherwise, it will not be $\varepsilon$-stable.
Due to the interacting queues problem, obtaining the exact sufficient and necessary condition for $\varepsilon$-stability, i.e., finding the critical arrival rate, is difficult.
In the following, we list several promising approaches.

\textbf{Sufficient Conditions} --
To obtain sufficient conditions for $\varepsilon$-stability, a dominant system can be considered, in which the transmission under consideration operates just like that in the original system. Other transmitters in the dominant system, when their queues are empty, will transmit ``dummy'' packets and continue to cause interference.
As a result, the queue lengths in the dominant system are larger than that in the original system, if the initial conditions of all queues in the two systems are the same.
Sufficient conditions for $\varepsilon$-stability of original system are obtained by analyzing the conditions for $\varepsilon$-stability of the dominant system.

\textbf{Necessary Conditions} --
We describe two methods to obtain two kinds of necessary conditions for $\varepsilon$-stability, which we name \emph{type I} and \emph{type II} necessary conditions.
To obtain \emph{type I necessary conditions}, we introduce a simplified system that just considers one nearest interferer.
A necessary condition for a queue in the original system to be stable is that the queue in the simplified system is stable because the interference in the simplified system is reduced.
To obtain the \emph{type II necessary conditions}, a modified favorable system can be introduced in which a packet at interfering transmitters is dropped if it is not scheduled or fails to be delivered.
Since the interference in the modified favorable system is smaller than that in the original
system, the necessary conditions for $\varepsilon$-stability of the original system can be obtained via deriving necessary conditions for $\varepsilon$-stability of the desired link in the modified favorable system.
Through introducing these two systems, whether an interfering transmitter is active or not is independently from the statuses of the queues. This way, the interacting queues become decoupled.

Consider a network model given by Fig. 1(c), where the nodes are modeled by a Poisson bipolar process with intensity $\lambda$ access points/m$^2$.
The time is slotted, and the packet arrival process is Bernoulli with arrival rate $\xi$.
The access probability for each link in each time slot is $p$. A retransmission will be conducted if a transmission fails in a certain time slot.
Based on the above approaches, we derive sufficient conditions as well as necessary conditions of $\varepsilon$-stability and relaxed them to closed-form  \cite{7486114}.
Figure \ref{fig:Com_p1} shows an example of the maximal arrival rates for sufficient conditions as well as necessary conditions when varying the access probability with random access, i.e., each link is scheduled independently with certain probability.
As the access probability approaches zero, packets in the modified favorable system are dropped with large probability. As the density of transmitters approaches zero, the interference is negligible. Therefore, applying the type I necessary conditions is a more appropriate choice than the type II necessary conditions in these cases.
When $\varepsilon\rightarrow0$, the type I necessary conditions become worse because the arrival rate may not be zero to achieve the strict stability ($\varepsilon=0$) of the simplified system, which is not consistent with the original system.
As a summary, the chosen of the appropriate type of necessary condition for different cases is in Table \ref{table:condition}.

\subsection*{Transmission Delay under Backlogged Assumption}
One way to bypass the interacting queues problem and analyze the delay is to assume that all nodes are fully backlogged, i.e., that the transmitters always have packets to deliver.
In this way, the service process at a transmitter is decoupled from the statuses of all other queues.
A meaningful and practically relevant metric under the backlogged assumption is the transmission delay, which is the duration to successfully deliver one packet.
The main component of the transmission delay is the retransmission delay and the waiting delay which are closely related to the number of retransmissions of a packet.
This type of delay, which ignores the queueing delay, is also called \emph{local delay} \cite{baccelli2010new}.

As discussed above, the delay is greatly affected by the interference correlation in the static ultradense networks, which may come from the correlated shadowing or fading. But more importantly, such interference correlation is caused by the spatial distribution of nodes as well as the MAC
mechanisms because they decide on the activity pattern of interfering transmitters, which, in turn, determines the spatiotemporal structure of the interference.

In static deployments, the transmission delay for different links in the extreme case without fading and MAC mechanism is either one time slot or infinite.
This is because when the realization of the point process modeling the nodes' distribution is good, a transmission will be always successful.
On the contrary, when the realization of the point process is bad, a transmission will be always failed, resulting infinite delay.
Thus, the events for successful transmissions in different time slots are completely correlated (one
successful transmission indicates successful transmissions in all time slots, and vice versa), resulting in infinite average transmission delay.

Figure \ref{fig:local delay} shows the mean and variance of the transmission delay with random access and backlogged nodes.
The mean transmission delay might become infinite under some system configurations --- a phenomenon named \emph{wireless contention phase transition}.
The variance reflects the delay jitter (delay fluctuation). For realtime applications like VoIP, a large variance of delay may cause a severe problem.
This approach has also been applied to analyze the transmission delay in heterogeneous cellular networks \cite{7247761}.

\subsection*{Single-hop Delay: Bounding Approaches}
The total single-hop delay is composed of the queueing delay and the
transmission delay.
In a static ultradense network, given the locations of transmitters and receivers, the
success probabilities for different links are different, resulting in different mean
delays for different links.
If we consider the mean delays of all links, a cumulative distribution function (cdf) of the mean delays of all queues can be obtained, which is a suitable metric to characterize the delay of the overall ultradense network.
Analytically, for ergodic point process models, the cdf obtained through the spatial statistics can be obtained by considering the typical queue.
However, obtaining the exact cdf is untractable due to the interacting queues problem. Several promising approaches to bound and approximate this cdf are described as follows.

\textbf{Lower bound} --
Considering the same dominant system introduced when deriving sufficient conditions for $\varepsilon$-stability,
the queue length for each link is larger than that in original system, leading to smaller SINR and larger delay. Therefore, the cdf obtained under such relaxation is a lower bound for the cdf of the mean
delays of all queues in the original system.

\textbf{Upper bound} --
Considering the same modified favorable system introduced when deriving type II necessary conditions, the interference is smaller than that in original system, leading to a smaller delay compared with that in the original system. Accordingly, the corresponding cdf is an upper bound for the cdf in the original system.

\textbf{Approximation} --
To approximate the cdf, all transmitters may be assumed to be busy independently with the same busy probability, which can be obtained by solving a fixed-point problem, as in \cite{stamatiou2014delay}.
The fixed-point problem is established by taking the busy probability of all interfering transmitters as a variable and expressing the busy probability of one desired link, which then equals the originally assumed busy probability of the interfering transmitters.
Having obtained the approximated busy probability, the cdf can then be approximately evaluated.

By applying the proposed bounding and approximating techniques, the service rate is decoupled from the statuses of all queues at the interfering transmitters and the analysis of the end-to-end delay becomes tractable.
Figure \ref{fig:Com_p} shows a comparison of the bounds for the cdf of the mean delay for different setups.

\section*{Conclusion and Future Research}
The coupling between traffic and network services becomes increasingly strong as ultradense networks become a reality.
This type of coupling gives rise to the interacting queues problem, which
is the key obstacle for delay analysis in ultradense networks.
This article proposed several promising approaches toward an understanding of the stability and delay issues in  ultradense networks.

Much work is still called for in this area, both on the fundamental theory to handle more sophisticated MAC protocols and on meaningful models to fit practical scenarios.
Some interesting aspects that need further investigation are as follows:
\begin{itemize}
\item Scheduling, which increases the complexity of the interacting queues, is a crucial mechanism that affects delay in the ultradense networks. More effective approaches should be investigated in order to analyze the impact of sophisticated scheduling mechanism.

\item Traditional traffic analyses either focus on modeling the spatial distribution of the traffic or modeling the temporal arrival process of packets. The methods discussed in this article are promising in jointly handling
    the spatiotemporal arrival of traffic.

\item Delay for more complicated yet realistic point process models, such as cluster processes and hard core processes, also needs to be explored. New approaches are also needed to obtain more accurate results for stability and delay distribution.
\end{itemize}

\bibliographystyle{IEEEtran}
\bibliography{123}

\newpage

\begin{figure}
\centering
\includegraphics[width=0.7\textwidth]{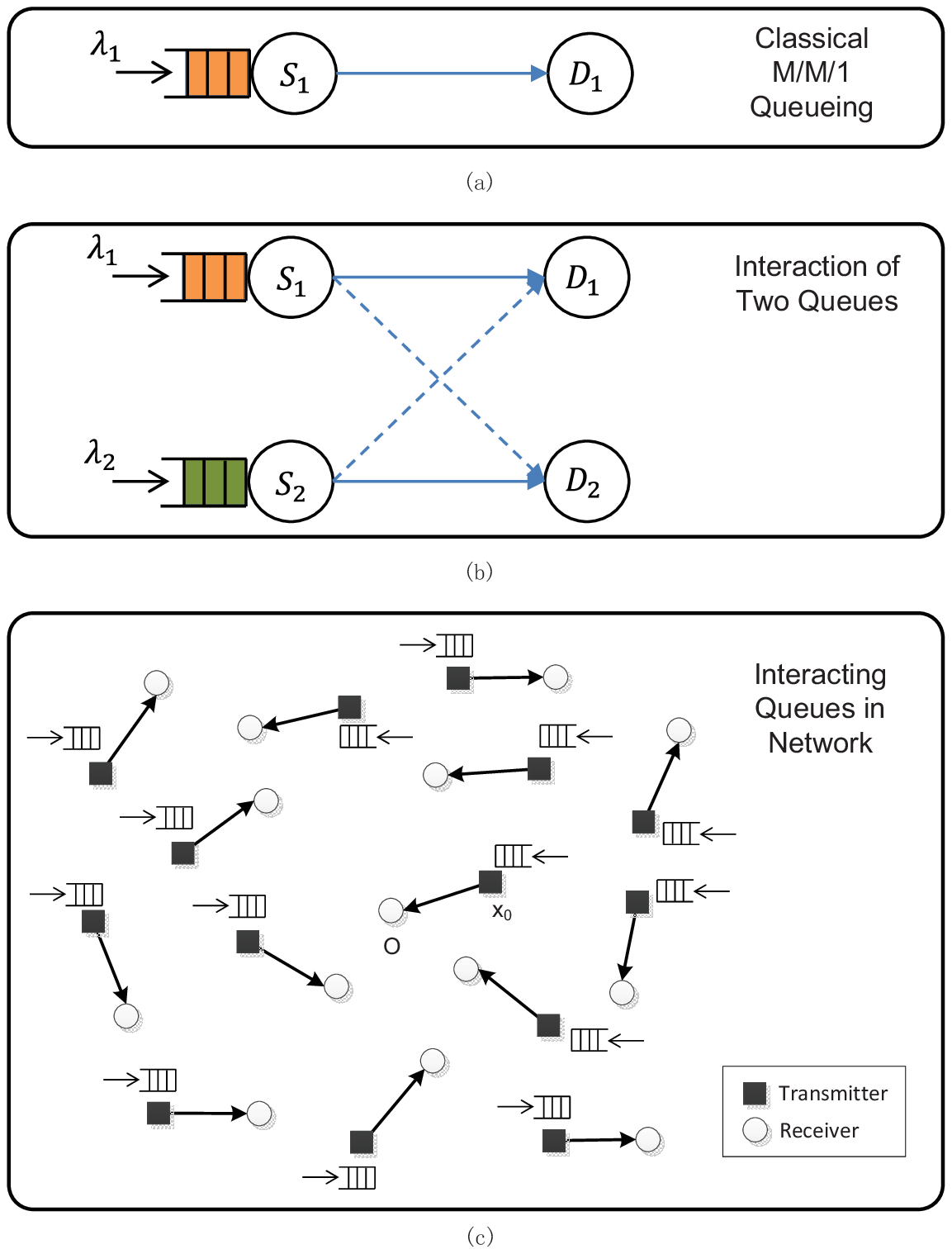}
\caption{(a) Classical M/M/1 queueing problem. (b) Interacting queues problem with two queues. (c) Interacting queues problem in ultradense networks.}
\label{fig:bipolar}
\end{figure}

%

\clearpage{}

\begin{figure}
\centering
\includegraphics[width=0.75\textwidth]{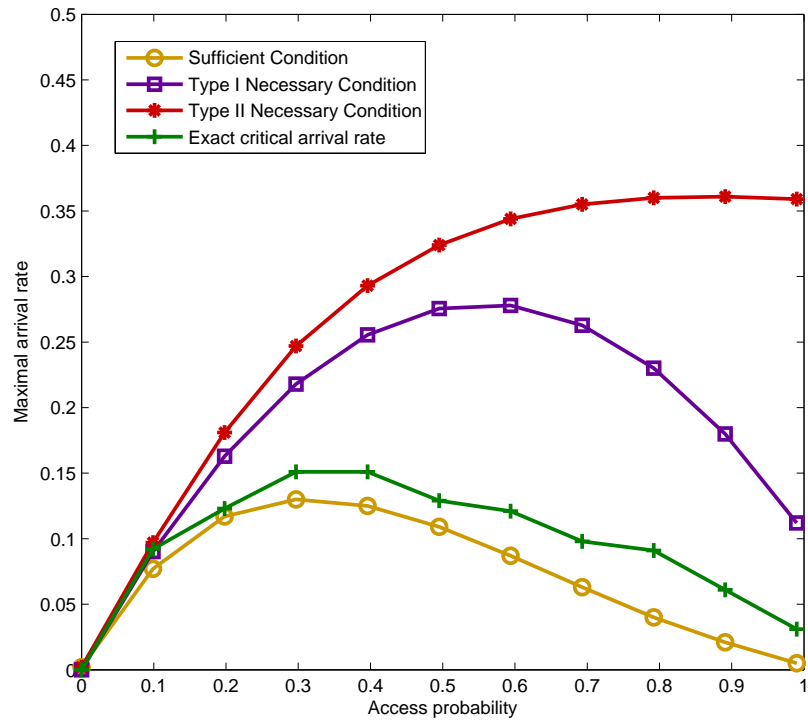}
\caption{Comparison of sufficient conditions and necessary conditions for $\varepsilon$-stability with $\varepsilon=0.1, \lambda=0.05$, and the distance between transmitter and receiver is $1$m.}
\label{fig:Com_p1}
\end{figure}

\clearpage{}

\begin{table}
\centering
\caption{Cases to choose type I or type II necessary conditions}
\label{table:condition}
\begin{tabular}{|c|c|}
\hline
\textbf{Some special cases} & \textbf{Type} \\
\hline
Parameter $\varepsilon$ for $\varepsilon$-stability approaches zero & Type II \\
\hline
Access probability approaches zero & Type I \\
\hline
Density of transmitters approaches zero  & Type I \\
\hline
SINR threshold $\theta$ approaches zero and access probability approaches one & Type II \\
\hline
Square of the desired link distance is much larger than reciprocal of the density of transmitters & Type II \\
\hline
\end{tabular}
\end{table}


\clearpage{}

\begin{figure}
\subfloat[]{\includegraphics[width=0.5\textwidth]{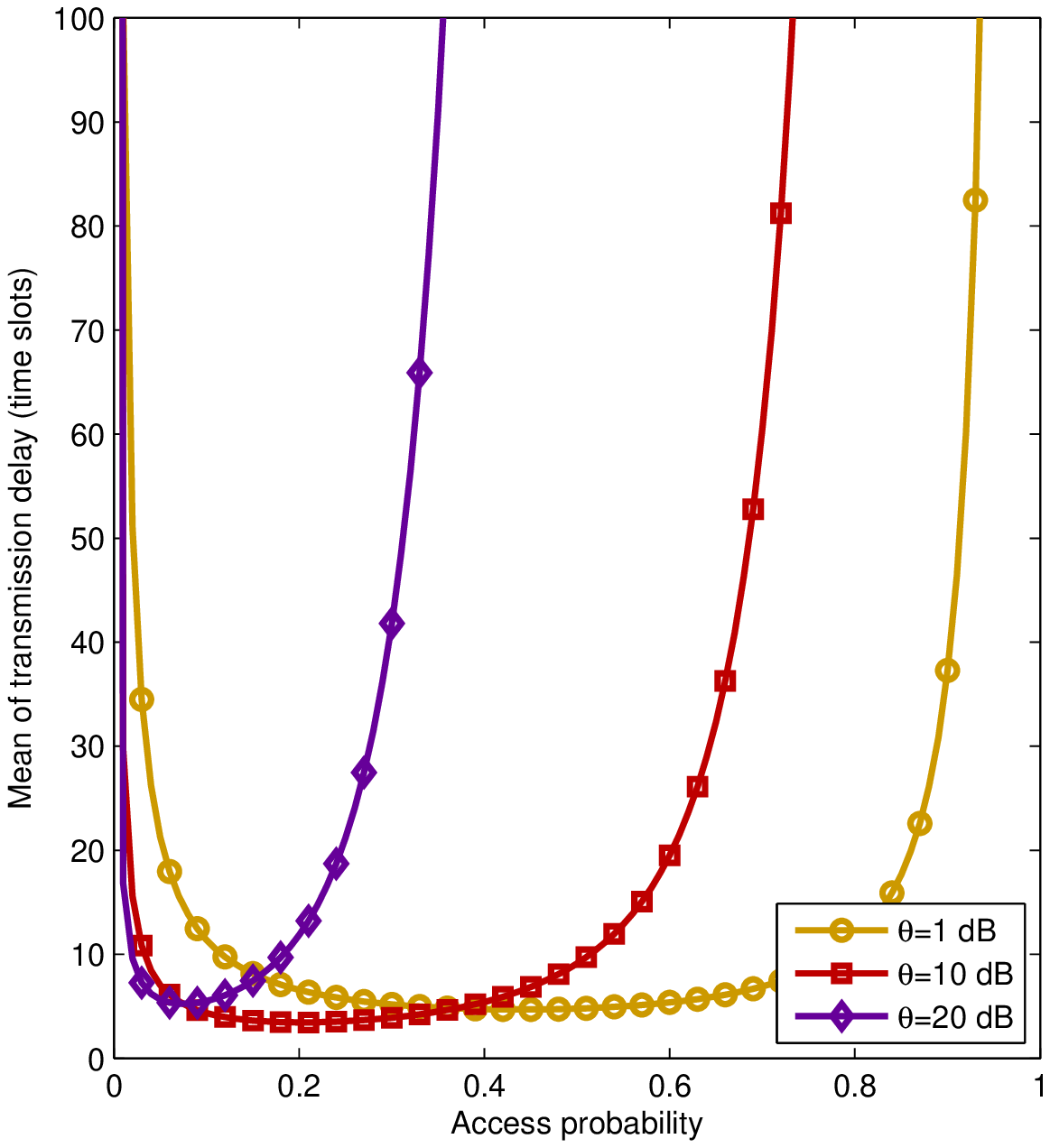}}\subfloat[]{\includegraphics[width=0.5\textwidth]{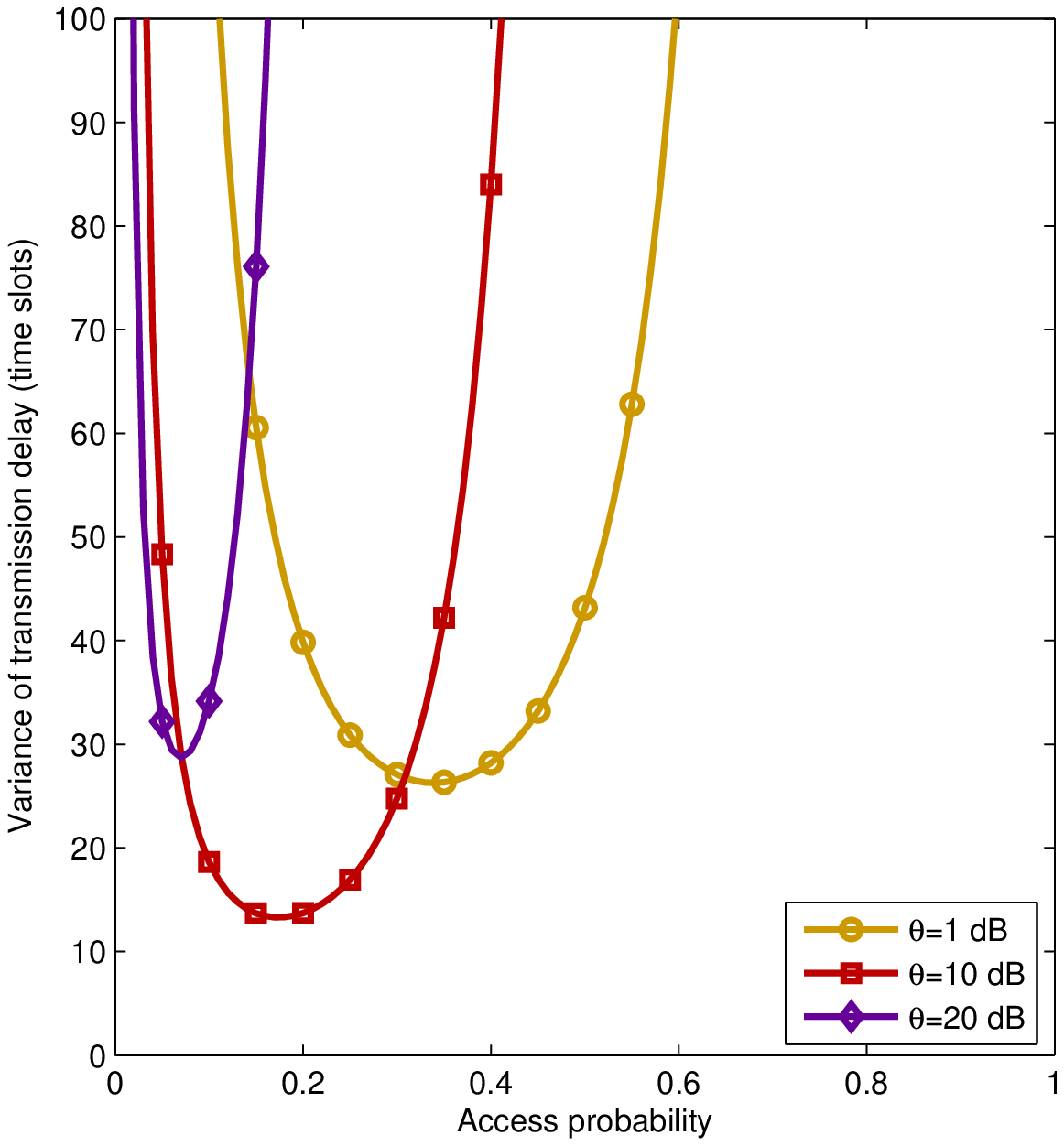}}\caption{\label{fig:local delay} Mean and variance of the transmission delay with random access and backlogged assumption. The distance between transmitter and receiver is $5$m, $\lambda=0.01$, and $\theta$ is the SINR threshold.}
\end{figure}

\clearpage{}

\begin{figure}
\centering
\includegraphics[width=0.85\textwidth]{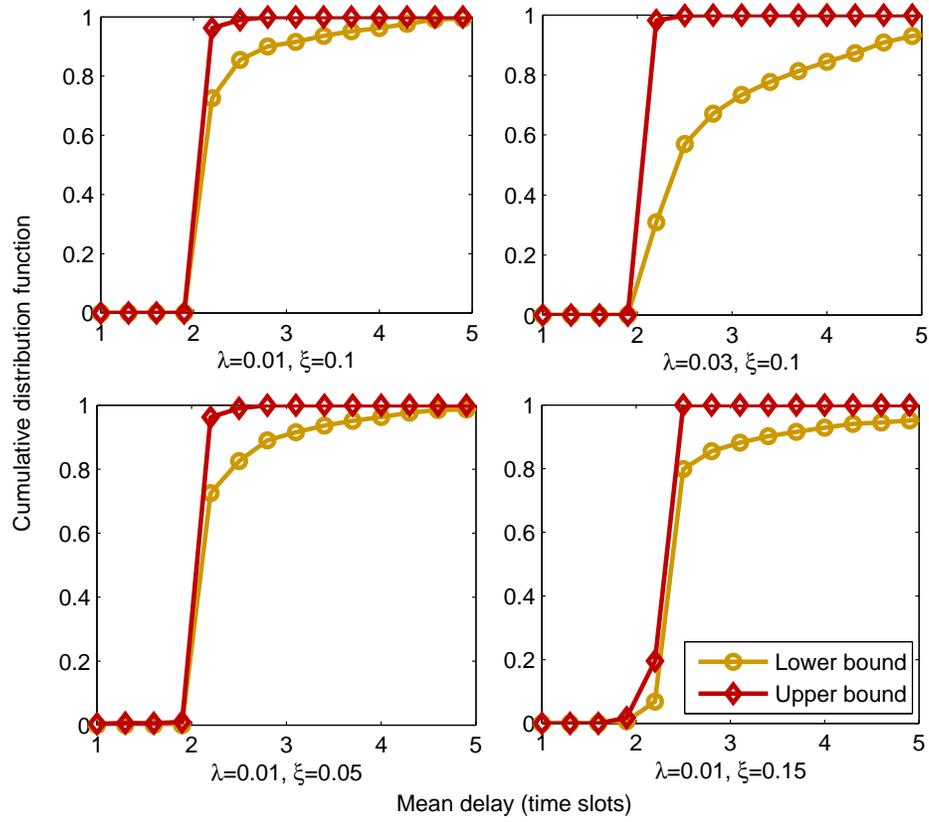}
\caption{Comparison of lower bound and upper bound for the cdf of
the mean delay with $p=0.5$, and the distance between transmitter and receiver is $1$m.}
\label{fig:Com_p}
\end{figure}


\clearpage{}

\begin{IEEEbiographynophoto}{Yi Zhong} (S'12-M'15) is an assistant professor with School of Electronic Information and Communications, Huazhong University of Science and Technology, Wuhan, China. He received his B.S. and Ph.D. degree from University of Science and Technology of China in 2010 and 2015 respectively. After that, he was a Postdoctoral Research Fellow with the Singapore University of Technology and Design. \end{IEEEbiographynophoto}

\begin{IEEEbiographynophoto}{Martin Haenggi} (S'95-M'99-SM'04-F'14) is the Frank M. Freimann Professor of electrical engineering at the University of Notre Dame, IN, USA. He received the Dr. sc .techn. (Ph.D.) degree from ETH Zurich in 1999. He is the Editor-in-Chief of the IEEE Transactions on Wireless Communications. He received the 2010 Best Tutorial Paper award, the 2017 Stephen O Rice Prize, and the 2017 Best Survey Paper award. \end{IEEEbiographynophoto}

\begin{IEEEbiographynophoto}{Fu-Chun Zheng} (M'95-SM'99)
obtained his PhD degree in Electrical Engineering from Edinburgh University, UK, in 1992, and has since held academic positions in Australia, UK, and China. He is currently with Harbin Institute of Technology, Shenzhen, China, and York University, York, UK. He was the executive TPC Chair for IEEE VTC 2016-S (www.ieeevtc.org/vtc2016spring).
\end{IEEEbiographynophoto}

\begin{IEEEbiographynophoto}{Wenyi Zhang} (S'00, M'07, SM'11) is with the faculty of the Department of Electronic Engineering and Information Science, University of Science and Technology of China. Prior to that, he has been with University of Southern California and Qualcomm Incorporated. He obtained Bachelor¡¯s degree in Automation at Tsinghua University in 2001, and Ph.D. degree in Electrical Engineering at the University of Notre Dame in 2006. \end{IEEEbiographynophoto}

\begin{IEEEbiographynophoto}{Tony Q.S. Quek}(S'98-M'08-SM'12) received the B.E.\ and M.E.\ degrees in Electrical and Electronics Engineering from Tokyo Institute of Technology. At MIT, he earned the Ph.D.\ in Electrical Engineering and Computer Science. Currently, he is a tenured Associate Professor with the Singapore University of Technology and Design. He also serves as the Associate Head of ISTD Pillar and the Deputy Director of the SUTD-ZJU IDEA. \end{IEEEbiographynophoto}

\begin{IEEEbiographynophoto}{Weili Nie} received the B.S. degree in electronics and information
engineering from Huazhong University of Science and Technology, China,
in 2012, and the M.S. degree in electrical engineering from Southeast
University, China, in 2015. He is currently pursuing his Ph.D. degree
in the ECE department at Rice University. \end{IEEEbiographynophoto}

\end{document}